# RNASeqR: an R package for automated two-group RNA-Seq analysis workflow

Kuan-Hao Chao, Yi-Wen Hsiao, Yi-Fang Lee, Chien-Yueh Lee, Liang-Chuan Lai, Mong-Hsun Tsai, Tzu-Pin Lu, and Eric Y. Chuang

**Abstract**—RNA-Seq analysis has revolutionized researchers' understanding of the transcriptome in biological research. Assessing the differences in transcriptomic profiles between tissue samples or patient groups enables researchers to explore the underlying biological impact of transcription. RNA-Seq analysis requires multiple processing steps and huge computational capabilities. There are many well-developed R packages for individual steps; however, there are few R/Bioconductor packages that integrate existing software tools into a comprehensive RNA-Seq analysis and provide fundamental end-to-end results in pure R environment so that researchers can quickly and easily get fundamental information in big sequencing data. To address this need, we have developed the open source R/Bioconductor package, RNASeqR. It allows users to run an automated RNA-Seq analysis with only six steps, producing essential tabular and graphical results for further biological interpretation. The features of RNASeqR include: six-step analysis, comprehensive visualization, background execution version, and the integration of both R and command-line software. RNASeqR provides fast, light-weight, and easy-to-run RNA-Seq analysis pipeline in pure R environment. It allows users to efficiently utilize popular software tools, including both R/Bioconductor and command-line tools, without predefining the resources or environments. RNASeqR is freely available for Linux and macOS operating systems from Bioconductor (https://bioconductor.org/packages/release/bioc/html/RNASeqR.html).

---

## 1 INTRODUCTION

RNA-Seq is a revolutionary approach with which to discover and investigate an entire transcriptome using next-generation sequencing (NGS) technologies [1]. The rapid advances in massive parallel RNA-Seq, as well as the decrease in cost and the development of various convenient analysis tools [2], have made RNA-Seq a widely used method for producing comprehensive transcriptional information. Typically, the main objective of this kind of transcriptome analysis is to identify genes that are differentially expressed under different conditions or in different tissues in order to gain an understanding of the physiological pathways associated with pathological conditions [1].

Many transcriptome analyses are case-control studies, i.e., observational studies focusing on an "exposure" (i.e., a risk factor or a medical treatment) which is correlated with a certain outcome [3]. A case-control experiment involves a group of samples with the outcome and a corresponding group without the outcome; it is retrospective in that it traces the different outcomes back to different exposures [3]. Examples of transcriptional case-control studies include the analysis of lesional (case) and normal (control) samples extracted via skin biopsies from *Homo sapiens* for increasing the understanding of psoriasis [4], and a comparison of the differential lipid metabolism gene expression between unparasitized *Aphids gossypii* and *Aphids gossypii* parasitized by *Lysiphlebia japonica* [5]. From these studies, we see that comparative RNA-Seq analysis is widely applicable and may play an important role in clinical or biological studies.

There are many well-developed algorithms and statistical methods for each analysis step of RNA-Seq, including aligners, transcript assemblers, and statistical methods for analyzing the differential expression of genes or transcripts in terms of either read counts or FPKM-based expression values [6]. However, this kind of piecemeal RNA-Seq analysis involves many software tools written in different programming languages for each processing step. It would be more efficient to integrate the individual analysis steps together into an automated workflow. The R programming language and its associated Bioconductor project have become popular and are widely used in many fields of biological research. While Bioconductor provides substantial R packages for certain parts of the overall pipeline of RNA-Seq analysis [7], there are few R packages for performing the analysis process in a comprehensive and reproducible way. Thus, an RNA-Seq analysis package within R will be beneficial for experimental or clinical exports.

RNASeqR was designed to offer an automated RNA-Seq workflow for running a transcriptome analysis pipeline in

- *K. Chao is with the Department of Electrical Engineering, National Taiwan University, Taipei, Taiwan. Email: b05901180@ntu.edu.tw.*
- *E. Chuang, Y. Hsiao and C. Lee is with the Bioinformatics and Biostatistics Core, Center of Genomic and Precision Medicine, National Taiwan University, Taipei, Taiwan. E-mail: y.w.hsiao9419@gmail.com, d00945006@ntu.edu.tw.*
- *E. Chuang, and Y. Lee is with the Graduate Institute of Biomedical Electronics and Bioinformatics, National Taiwan University, Taipei, Taiwan. Email: r05945010@ntu.edu.tw.*
- *L. Lai is with the Graduate Institute of Physiology, National Taiwan University, Taipei, Taiwan. E-mail: llai@ntu.edu.tw.*
- *M. Tsai is with the Institute of Biotechnology, National Taiwan University, Taipei, Taiwan. E-mail: motion@ntu.edu.tw.*
- *T. Lu is with the Institute of Epidemiology and Preventive Medicine, Department of Public Health, National Taiwan University, Taipei, Taiwan. Email: tplu@ntu.edu.tw.*

six steps with the support of comprehensive visualization, the integration of command-line tools in the R environment, and the option for functions operating either in an R-interactive version or in a background version.

## 2 METHODS

RNASeqR is an open-source Bioconductor package for analyzing two-group RNA-Seq data with at least three replicates in each group. It is designed to simplify the analysis procedure by implementing a standardized workflow [8], which provides reproducible and trustworthy results in the R environment. RNASeqR is available on Linux and macOS operating systems, with support for integrating the command-line interface tools HISAT2 [9], StringTie [10], and Gffcompare into the R environment, and is able to be run in an interactive R shell (R-interactive version) or in the background (background version). By running well-constructed functions in RNASeqR, the results files created from each step, in either tabular or graphical format, are kept in a well-defined file structure.

### 2.1 Workflow Design

The RNASeqR workflow (Fig. 1) is designed to conduct a two-group RNA-Seq data analysis in only six steps, the main design concept being that each step is implemented as an R function. This six-step procedure includes: (1) the creation of an S4 *RNASeqRParam* object, (2) the setup of the environment, (3) read quality assessment among all samples, (4) read alignment and quantification, (5) gene-level differential analyses, and (6) functional analysis. Users have to run these functions in order, as some steps will process only when the input file generated from the previous step exists. Before running subsequent steps, users must make sure the current step has finished successfully, whether in the R-interactive version or the background version. Some steps take about 6 hours to finish; therefore, it is highly recommended to run the background version. If the program terminates accidentally, users do not need to go into the project folder and remove any files. Simply rerun the function and everything will be redone. The only exception is the third step, Quality Assessment, which is an optional step that users can run any time they want.

## 3 RESULTS

To illustrate the utility of RNASeqR, the package workflow was applied to a set of raw RNA-Seq data. Six samples of *Saccharomyces cerevisiae* RNA were selected from NCBI's Sequence Read Archive (SRA) [11]: entries SRR3396381, SRR3396382, SRR3396384, SRR3396385, SRR3396386, and SRR3396387 in BioProject PRJNA318684. Suitable reference genome and gene annotation files in Ensembl version R64-1-1 were downloaded from the iGenomes Illumina support site (https://support.illumina.com/sequencing/sequencing_software/igenome.html). Three samples (SRR3396381, SRR3396382, SRR3396384) were treated with 0.03 μg/mL amphotericin B and are referred to as the case group, while the remaining samples (SRR3396385, SRR3396386, SRR3396387) did not receive any treatment and were the control group. Both sets of yeast cultures were grown in ID20 medium (inhibition of cell growth by 20%) for 60 minutes [12].

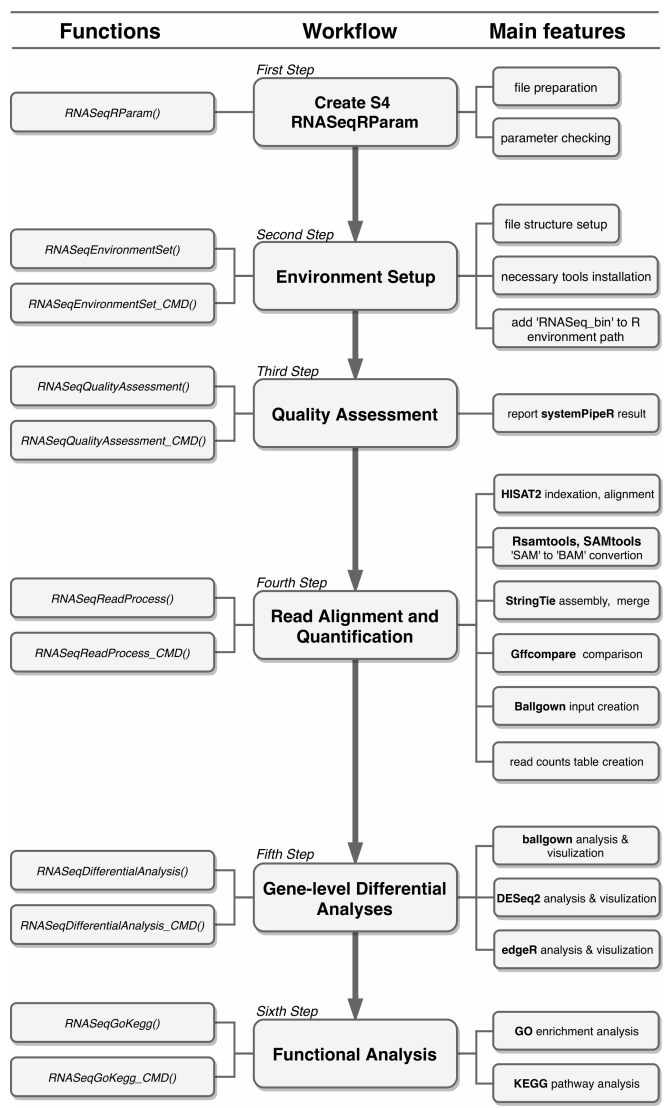

Fig. 1. Overview of the full workflow of RNASeqR. There are three parts in this figure: the main workflow (middle), the functions involved in each step (on the left), and the substeps that can be performed within each function (on the right). Aside from the function *S4 RNASeqRParam*, the remaining functions can be conducted in either the R-interactive version or the background version (by adding the *CMD* suffix). After correctly running the functions in order, a two-group comparative RNA-Seq data analysis will be done.

### 3.1 *RNASeqRParam* Object Creation

A new S4 class named *RNASeqRParam* was designed to contain all of the essential data required for transcriptome analysis. Users have to prepare an *input_files* directory containing raw files formatted in accordance with the file structure regulation (Fig. 2a). The consistency between user input parameters of the constructor function *RNASeqRParam()* and raw files in the *input_files* directory will be thoroughly checked, and any mismatch will cause this step to fail. The successfully created *RNASeqRParam* S4 object is used as the input parameter in the subsequent analysis

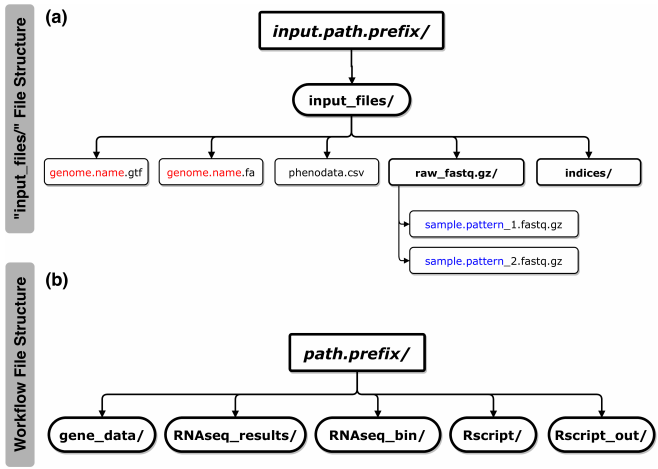

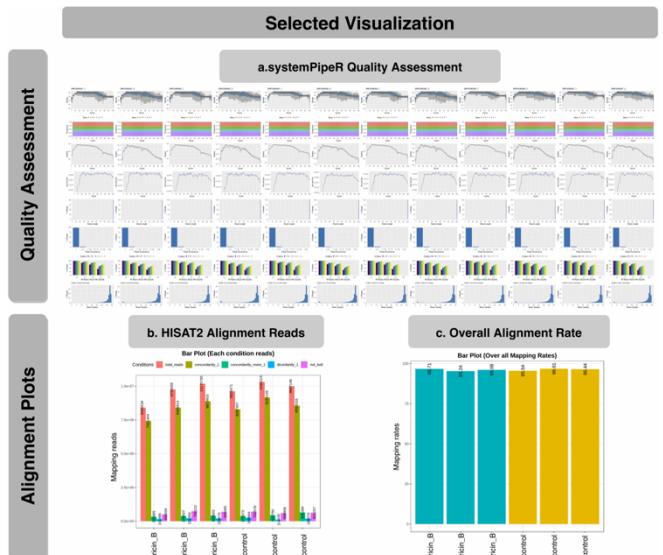

Fig. 2. a The structure of the *input_files* directory. *input.path.prefix* is the parameter of the absolute path in which the *input_files* directory is located. Under the *input_files* directory, there are several necessary files that users have to provide in order to pass the checking procedure of the RNASeqRParam constructor function: (i) *genome.name.gtf*, (ii) *genome.name.fa*, (iii) *phenodata.csv*, (iv) the *raw_fastq.gz* directory containing raw paired-end fastq.gz files of all samples and (v) the *indices* directory (optional) for storing HISAT2 indexes. The *genome.name* and *sample.pattern* files marked in red and blue, respectively, are the constructor function parameter values. b The structure of the RNASeqR root directory. *path.prefix* is the absolute path of the RNASeqR root directory. Five main directories are created in the second step, Environment Setup. (i) *gene_data* contains symbolic links to gtf, fa, fastq.gz, and ht2 files and intermediate files in bam, sam, etc.; (ii) *RNAseq_results* contains alignment results, differential gene expression results, and functional results in tabular, png or pdf format; (iii) *RNAseq_bin* contains HISAT2, StringTie, and Gffcompare binaries in compressed and decompressed forms; (iv) *Rscript* contains R scripts that are created by running background version functions (with the CMD suffix); and (v) *Rscript_out* contains Rout log files created by running background version functions.

Fig. 3. Selected visual results of the Quality Assessment and Read Alignment steps. a Quality assessment results reported by the systemPipeR Bioconductor package providing sequence quality score distribution, GC content, read length distribution, etc. b HISAT2 alignment results for each sample: (i) total reads, (ii) aligned concordantly exactly 1 time, (iii) aligned concordantly >1 time, (iv) aligned discordantly exactly 1 time, and (v) reads not aligned. c The overall HISAT2 alignment rate of each sample. Blue and yellow represent two distinct groups.

functions, which can simplify the whole RNA-Seq data analysis process.

### 3.2 Environment Setup
By running the function for this step, either *RNASeqEnvironmentSet()* or *RNASeqEnvironmentSet_CMD()*, the basic user-defined file structure (Fig. 2b) is created, and the necessary third-party command-line interface software programs used in this workflow, e.g., HISAT2, StringTie, and Gffcompare, are downloaded automatically. The *RNASeq_bin* directory containing these decompressed binaries is automatically added to the R environment path variable.

### 3.3 Quality Assessment
Quality assessment of raw reads is an important step in RNA-Seq data analysis, leading to high quality outcomes. This mainly includes calculating the sequence quality score and GC content, and checking for the presence of the adapter sequence [13]. After running *RNASeqQualityAssessment()* or *RNASeqQualityAssessment_CMD()*, the quality report of all samples is created by systemPipeR [14] and kept in the *RNASeq_results* directory (Fig. 3a). The assigned score of each base call, also called the Phred score, refers to the probability of a base being incorrect [15]. Higher value indicate a higher chance that a call is correct. Generally, it is recommended that the average score of each sample should at least reach Q30, which means each base call has a 1 in 1000 probability of being incorrect (i.e., 99.9% base call accuracy) [16]. In RNASeqR, a sample with an average quality score below Q30 is suggested to be omitted from the subsequent analyses. In the case of our *S. cerevisiae* RNA-Seq experiment, the average score of all samples was over Q30, indicating the quality of this set of sequencing data was sufficient to proceed with the alignment step (Fig. 3a).

### 3.4 Read Alignment and Quantification
Reads are typically mapped to an annotated genome, and then we quantify the number of reads that are assigned to a specific gene region based on the comprehensive gene annotation file. A correct reference genome sequence and accurate gene annotation lead to accurate estimation of gene expression. This essential and all-in-one step involves reference indexing (HISAT2), read alignment (HISAT2), SAM-to-BAM file conversion (Samtools), transcript assembly (StringTie), transcript comparison (Gffcompare), ballgown input creation, and read count table creation (StringTie). All of these substeps are completed after running *RNASeqReadProcess()* or *RNASeqReadProcess_CMD()*. However, this step takes considerable time to finish as it contains several substeps. Taking the *S. cerevisiae* RNA-Seq experiment as an example, six FASTQ files, each with 8.4-10.3 million reads, were processed using 16 CPU cores and 160 GB of RAM; this consumed 14,531.328 seconds for the user CPU time and 4,184.594 seconds for the system CPU time (>5 hours in total). Therefore, *RNASeqReadProcess_CMD()*, which creates an R script and processes in the background, is highly recommended for this step. Alternatively, under the premise that certain substeps are already successfully finished, users could skip them by setting the corresponding function parameter in order to speed up the whole process.

The percentage of mapped reads is an important parameter for the overall alignment accuracy. For a model organism, it is recommended that 70-90% of RNA-Seq data should map to its corresponding genome [17]. To check mapping quality, in our package, different mapping types and overall alignment rates of each sample are summarized as tabular output and visualized using bar charts (Fig. 3b & 3c). Overall, the mapping rates of all the *S. cerevisiae* samples were over 95%, illustrating that both the library of sequencing reads and the reference genome for read alignment are highly accurate.

### 3.5 Gene-level Differential Analyses

The most common application of RNA-Seq is to identify gene expression changes between experimental conditions, elucidating possible biological mechanisms behind such experimental designs. In this step, three Bioconductor packages, supporting the detection of differentially expressed genes (DEGs), are applied: ballgown [18], DESeq2 [19], and edgeR [20, 21]. The default thresholds of the p-value and $\log_2$ fold change for DEGs are 0.05 and 1, but they can be manually modified for specific needs. RNASeqR provides comprehensive outputs in both tabular and graphical format by running *RNASeqDifferentialAnalysis()* or *RNASeqDifferentialAnalysis_CMD()*. All visual plots created during this process are stored separately in the respective image directories under *ballgown_analysis*, *DESeq2_analysis*, or *edgeR_analysis*.

In each directory, images are divided into two categories: pre-differential expression (*preDE* directory) and differential expression (*DE* directory). Prior to differential expression analysis, the distribution of gene expression of each sample in terms of normalized counts or FPKM value is displayed as frequency plots and box plots (Fig. 4a & 4b) using ggplot2 [22] in order to check whether the normalization approach is suitable for each dataset. For example, in our *S. cerevisiae* case, both plots showed a consistent distribution profile when applying the proper normalization. The pairwise correlations of expression profiles between all samples is visualized as a heatmap (Fig. 4c) using corrplot [23] and shows how the samples are related to each other. Samples with higher similarity to one another will be nearer each other. This helps users to further identify outlier samples whose correlation with all other samples is low. Likewise, principal component analysis (PCA), a multivariate analysis technique, is used to represent the whole gene expression profile of each sample as a single value [24]. The two-dimensional PCA plot (Fig. 4d) is generated using FactoMineR/factoextra [25] and ggplot2 R packages for exploring within- and between-group clustering. It allows users to identify how sample classes separate and exclude the outlier. In the two-group *S. cerevisiae* RNA-Seq case, both the correlation plot and the PCA plot (Fig. 4c & d) revealed that the clustering of case versus control samples was good, without any outliers that should be dropped out. Moreover, RNASeqR also provides package-specific plots for each DE analysis package, such as the distribution of transcript length produced by ballgown, shown in Fig. 4e. After differential expression analysis using three approaches, four common plots are generated in each DE folder: a PCA plot, a mean average plot, a volcano plot, and a heatmap. The PCA plot, at this stage, illustrates whether expression of selected DEGs successfully separates sample classes as well (Fig. 4f). In our *S. cerevisiae* example, the variance explained by the first principal component increases from 50.27% to 90.5%, indicating that the expression of those selected DEGs is sufficient to distinguish the difference between conditions. The mean average plot of the $\log_2$ fold change for all detected genes and the volcano plot comparing the statistical significance and the fold change for each gene are generated using ggplot2 (Fig. 4g & 4h). Significant DEGs are highlighted in both plots to visualize whether the proportion of DEGs across all detected genes under default cutoffs is appropriate. To further visualize the expression of DEGs in the *S. cerevisiae* samples, a heatmap is also generated using the pheatmap [26] R package (Fig. 4i). It allows the user to identify up-regulated and down-regulated genes in the case group compared with the control.

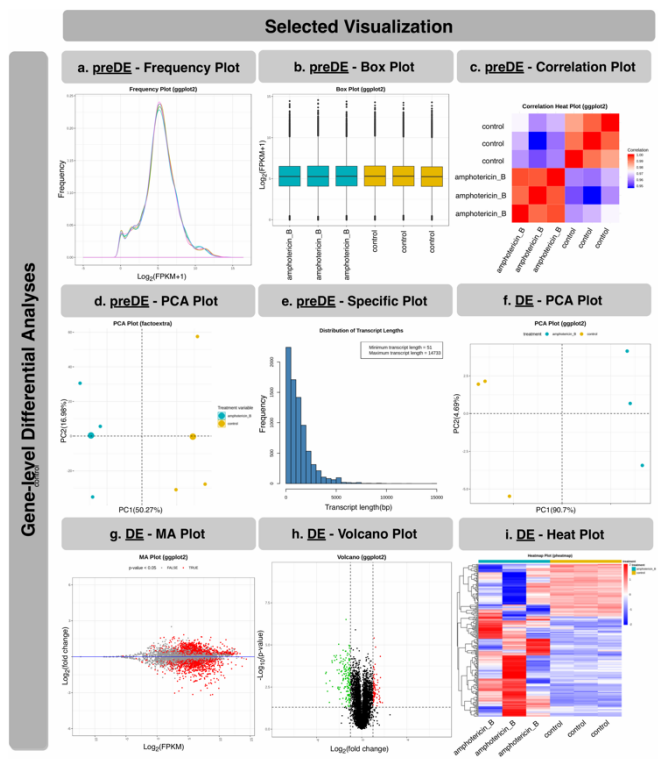

Fig. 4. Selected visual results of gene-level differential analyses. a Frequency plot of normalized gene expression. Each curve represents one sample. b Box plot of normalized gene expression. c Principal Component Analysis (PCA) plot of all samples. The x-axis is the most dominant principal component (PC1); the Y-axis is the second most dominant principal component (PC2). d Correlation heatmap plot with the depth of color representative of correlation between samples. Red and blue represent the maximum and minimum correlation values, respectively. e Distribution plot of transcript length from the ballgown package. Package-specific plots are available for each differential analysis tool. f PCA plot of all samples using the expression values of DEGs. g Mean average (MA) plot. Each dot corresponds to a gene. Genes with a p-value smaller than 0.05 are marked in red. h Volcano plot. Each dot represents a DEG. Down-regulated and up-regulated genes, based on default filter criteria, are marked in green and red, respectively. i Sample-feature (Gene) heatmaps using k-means clustering on gene expression. The log2-nomalized (FPMK+1) values of all the genes were used. The two sample classes (i.e., case and control) are indicated in blue and yellow, respectively.

### 3.6 Functional Analysis

It is important to understand the biological impact of DEGs involved in the condition of interest. After gene-level differential analyses, users can run *RNASeqGoKegg()* or *RNASeqGoKegg_CMD()* to determine the enriched biological processes/pathways of DEGs found by the analysis using ballgown, DESeq2, and edgeR. The clusterProfiler package [27] used in this step provides two main features: Gene Ontology (GO) [28, 29] analysis and Kyoto Encyclopedia of Genes and Genomes (KEGG) [30, 31] pathway analysis. Both features are integrated into RNASeqR. For GO analysis, classification and over-representation results are presented in terms of three aspects of GO: molecular function, biological process, and cellular component. In the statistical output from DESeq2, for instance, the bar chart for the GO classification of DEGs (Fig. 5a) shows the top 15 biological processes involved in amphotericin B treatment of *S. cerevisiae*. The dot plot (Fig. 5b) and bar chart (Fig. 5c) of GO over-representation illustrate that the nucleolus and preribosome are the two significantly involved cellular components with smallest p-values and highest counts. For KEGG pathway analysis, enriched pathways are stored in csv format and the five most significant pathways are visualized by pathview [32], a Bioconductor package (Fig. 5d). Furthermore, the URL of the pathway is also stored in txt format on the KEGG website (Fig. 5e). Once this step is successfully finished, a comprehensive RNA-Seq data analysis using RNASeqR is complete.

### 3.6 More Results from Published Data

Additional two-group RNA-Seq data analysis examples are available. RNA-Seq analysis results for Gene Expression Omnibus (GEO) [33, 34] entries GSE100075 [35] and GSE50760 [36] from *Homo sapiens* using RNASeqR are presented in Additional files 5, 6 and 7.

## 4 DISCUSSION

There are several reasons that R was chosen as the development platform for RNASeqR. First, R is currently one of the most popular languages for statistical data analysis in bioinformatics and data science. Second, R is an open source and open development language providing simple data manipulation methods and powerful visualization utilities for developers [37]. Third, Bioconductor provides more than 1600 high-quality R software packages for statistical analysis and genomic data processing in each step of a high-throughput method. Moreover, there are also many annotation, experiment, and workflow packages on Bioconductor, which makes the R environment developer-friendly [38, 39]. Fourth, the R language is supported by Bioconda. Bioconda is a bioinformatics channel hosted on GitHub that turns recipes into conda packages, and is maintained by the bioinformatics community [40]. Bioconda provides a nice platform for packages supported by both Linux and macOS operating systems, which makes the installation process much easier.

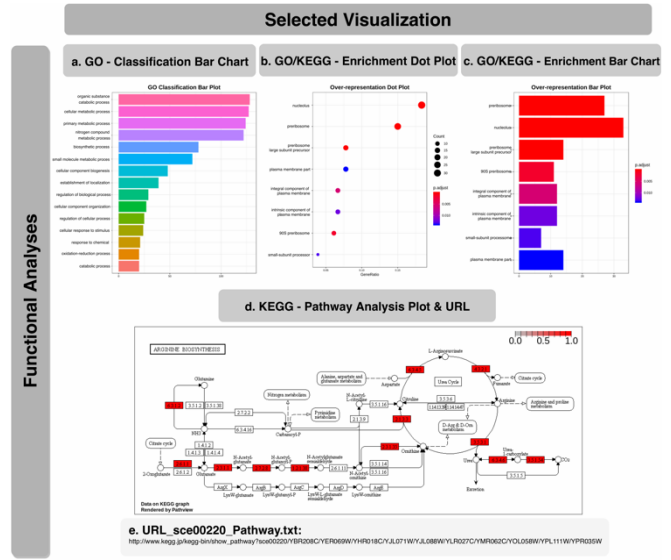

Fig. 5. Selected visual results of functional analyses. a The top 15 GO classification results in terms of molecular function, cellular component and biological process are visualized using a bar graph. Enrichment results of GO analysis and KEGG pathway analysis are presented using both a bar graph (b) and a dot plot (c). d Visualization of significantly enriched pathway IDs found by KEGG pathway analysis with the pathview Bioconductor package. Pathway ID sce0020 is illustrated as an example and the corresponding URL on the KEGG website is provided in (e).

### 4.1 Reasons to Implement this Pipeline

RNASeqR provides a fast way for researchers to quickly get overview of sequencing data in pure R environment and its six-step approach is the most convenient and easiest method among all current RNA-Seq pipeline tools; therefore, it would assist clinical researchers without significant computational background to grasp fundamental RNA-Seq results easily.

In RNASeqR, the new tuxedo pipeline published in *Nature Protocols* in 2016 [8] is fully implemented in the R environment, including extra features such as quality assessment, differential expression analysis, and functional analysis. Each tool in RNASeqR is carefully selected. In this pipeline, both command-line interface tools and R packages are used. Conducting an RNA-Seq analysis piecemeal in separate environments using similar repeat commands for each sample would be inefficient and monotonous for users. Therefore, RNASeqR is designed to integrate command-line interface tools into an R package and automate each step. The Following are the reasons for selected tools.

HISAT2 is the alignment tool used in RNASeqR, and it was chosen because of the following advantages. (i) HISAT2 is a highly memory-efficient system using the Burrows-Wheeler transform and the Ferragina-Manzini index. Moreover, it supports genomes of any size, which makes this pipeline widely applicable. (ii) Its processing speed is faster compared with OLego [41], STAR [42], and TopHat2 [43], and therefore it speeds up the whole RNASeqR analysis procedure. (iii) Its sensitivity and precision are higher than other alignment tools. The real data results of 20 million simulated 100-bp reads and small anchor reads revealed that HISAT2 series tools have the greatest accuracy [9].

Regarding transcript assembly and quantification, StringTie and Cufflinks [44] are the two most popular tools in RNA-Seq analysis. StringTie was chosen for use in RNASeqR for the following reasons. (i) StringTie has better performance on identifying the dominant transcript for a gene locus. For example, the number of transcripts only detected by StringTie is higher than those only detected by Cufflinks; therefore, it is better in terms of isoform prediction. (ii) StringTie is superior at reconstructing genes with low abundance, more exons, or multiple isoforms. (iii) StringTie runs faster than Cufflinks and Traph [45]. The lower processing time of StringTie speeds up the RNASeqR pipeline [10].

Ballgown [46] is a statistical package for estimating differentially expressed genes, transcripts, or exons. The benefit of ballgown is that it is highly compatible with the output from StringTie, which is suitable for the RNASeqR workflow. Moreover, two popular Bioconductor tools for count-based differential gene expression analysis, DESeq2 and edgeR, are also implemented in this pipeline, providing results based on different statistical methods.

### 4.2 R-interactive version and background version

An important feature in RNASeqR is that it provides two options for running the workflow. Aside from the first step, there are two types of functions for the rest of the steps: R-interactive version and background version. The first one, R-interactive version, can be called and executed in the R shell just like normal functions. However, some functions take a great deal of time and are not practical to run in the R shell. With real 10 versus 10 (20 raw fastq.gz files, roughly 3 GB per file) whole human RNA-Seq data, for example, it took 206,755.42 seconds for the user CPU time and 148,708.59 seconds for the system CPU time under 160 GB of RAM in 16 threads (almost 100 hours total). Therefore, the background version function, which has the _CMD suffix, is preferable. By calling this function, a small script will be created and run in the background with *nohup R CMD*. The latest progress of the function will be reported in the Rout file of the *Rscript_out* directory. Users should ensure the function is finished successfully before running subsequent functions.

### 4.3 Well-defined File Structure

Because RNASeqR has its own well-established file structure, users only need to run functions in order and do not need to manually organize the file structure by themselves. The root file structure of RNASeqR (Fig. 2b) will be created in the second step, Environment Setup, and all files created in each subsequent step will be stored properly. For example, alignments, such as bam and sam files, and their statistical results will be stored in their corresponding directories; this facilitates the processing of subsequent analyses based on these preliminary files.

### 4.4 Comprehensive Visualization

Another main feature of RNASeqR is that it provides comprehensive visualized plots for users to further interpret the results. In Gene-level Differential Analysis (the fifth step), the results of ballgown, DESeq2, and edgeR will be created, and users can manually set the function parameters to skip any differential analysis tool. For the functional analyses, GO and KEGG results are visualized in bar graphs and dot plots. The high-quality figures generated in each step are all in png format at 300 dpi.

### 4.5 Command-line Software Integration

RNASeqR integrates many command-line tools, e.g., HISAT2, StringTie, and Gffcompare, into the R environment. Users do not need to download these tools on their own. Instead, all compressed binaries of these tools will be downloaded automatically during the second step, Environment Setup, based on the operating system that is detected on the workstation. Subsequently, they will be decompressed and stored in the *RNAseq_bin* directory, which will be added to the R environment path. The shell commands of these tools are available by running the *system2()* function in the R environment. Although there are so many substeps in RNA-Seq analysis, the only thing that users need to do is to call the *RNASeqEnvironmentSet()* or *RNASeqEnvironmentSet_CMD()* function and all the relevant tools will be set up accordingly.

### 4.6 Ease of Use

RNASeqR is an open source tool available on both Bioconductor and Bioconda, providing users with simple installation methods. Moreover, the latest version can also be downloaded directly through GitHub. All these methods allow users to install all dependent R packages with only a single command. Running RNASeqR only requires the fundamentals of the Linux operating system and R language; that is, users only need to run simple R functions in order and check the results files to make sure the process is completely finished for conducting a basic comparative RNA-Seq data analysis.

### 4.7 Additional experiment data package

Each Bioconductor package needs a comprehensive vignette to demonstrate the usage of each function. RNASeqRData, an additional Bioconductor experiment data package, was created for vignette demonstration of the RNASeqR software package. These RNA-Seq sample data were extracted from the data used in the case study. The original fastq.gz files are about 800 MB, which is too large for a vignette demonstration; therefore, to reduce the size of the files but keep as many differentially expressed genes as possible, only the reads aligned to the region from bases 0 to 100,000 on chromosome XV were extracted.

### 4.8 Comparison to Other Tools

There are several tools designed for NGS analysis. Galaxy [47], Taverna [48], Snakemake [49], Nextflow [50], CIPHER [51], and bcbio-nextgen [52] all provide NGS data analysis infrastructure in various computer language environments. For example, tools such as VIPER [2], TRAPLINE [53], HppRNA [54], and QuickRNASeq [55] are RNA-Seq pipelines based on the Snakemake framework, combining tools in numerous languages including R, Python, Perl, C++, or Java. Although there are many frameworks for RNA-Seq data analysis, only a few of them are able to run end-to-end RNA-Seq data analysis in a pure R environment.

RNASeqR is implemented as a Bioconductor R package. Bioconductor is a central repository storing bioinformatics-related R packages, which has the following advantages. (i) It resolves dependencies between packages. Installing packages manually might have version incompatibility problems. (ii) It provides reliable code for all packages. Software tools must pass technical review before being accepted, and developers have to provide a comprehensive vignette with runnable example code. (iii) Packages that are out of maintenance are removed regularly.

Most Bioconductor packages are only designed for a specific analysis step or part of the overall pipeline. Take QuasR [56] as an example; it provides a workflow from raw sequence reads to quality assessment and quantification of genomic regions, which are the initial steps of a few NGS pipelines, such as RNA-Seq and Bis-Seq. However, it lacks an interface with external third-party command-line tools and downstream analyses like differential gene expression and pathway analysis. Although some packages provide a comprehensive framework integrating external command-line tools into the R environment for a complete NGS pipeline, these still involve many manual operations for users.

For example, systemPipeR [14] provides a flexible framework for many analysis pipelines, such as RNA-Seq, CHIP-Seq, Ribo-Seq, and VAR-Seq, with external command-line tools in the R environment. However, the whole analysis pipeline is not completely automated. Users still have to run complex downstream analyses manually using other Bioconductor packages. In systemPipeR RNA-Seq workflow document in Bioconductor, users have to write fifty more lines of code manually in order to do downstream analysis. In contrast, only two functions are needed in RNASeqR in order to finish differential expression, functional analysis and get comprehensive visualization. Furthermore, RNASeqR provides both R-version and background-version to process commands.

Similarly, bcbioRNASeq [57], another R package for comprehensive RNA-Seq data analysis, takes bcbio [52] output as input and provides functions for data visualization and downstream analysis. However, users still have to run bcbio and bcbioRNASeq separately and manually with several lines of codes. Unlike these packages, RNASeqR not only provides a comprehensive RNA-Seq data analysis but also focuses on one specific pipeline which has been recently reported as the most efficient and widely-used workflow [8]. Furthermore, RNASeqR largely simplifies RNA-Seq analysis, starting from quality assessment to functional analysis, into just six functions and provides a large number of interpretable plots. Therefore, our package allows clinical researchers to speed up the process of RNA-Seq data analysis, retaining high accuracy and providing graphical results for further biological interpretation.

## 5 CONCLUSION

RNASeqR is a new Bioconductor package providing a six-step automated workflow for two-group comparative RNA-Seq analysis. The core design concept of RNASeqR is that each RNA-Seq analysis step is implemented as an R function in the package, and thus users can perform RNA-Seq analysis instinctively. The main features of RNASeqR include: (i) flexible function options in both an R-interactive version and a background version, (ii) comprehensive visualization, (iii) integration of command-line tools into the R environment, (iv) a well-defined file structure, and (v) ease of use. We believe that RNASeqR will assist clinical researchers without significant computational background to obtain useful information from RNA-Seq data in an easy, efficient and reproducible manner.

## ACKNOWLEDGMENT

This work has been supported in part by the Center of Genomic and Precision Medicine, National Taiwan University, Taiwan [106R8400]; and the Center for Biotechnology, National Taiwan University, Taiwan [GTZ300].
We thank Melissa Stauffer, Ph. D., for editing the manuscript.

## REFERENCES

[1] Z. Wang, M. Gerstein, and M. Snyder, "RNA-Seq: a revolutionary tool for transcriptomics," *Nat Rev Genet,* vol. 10, no. 1, pp. 57-63, Jan, 2009.

[2] M. Cornwell, M. Vangala, L. Taing, Z. Herbert, J. Koster, B. Li, H. Sun, T. Li, J. Zhang, X. Qiu, M. Pun, R. Jeselsohn, M. Brown, X. S. Liu, and H. W. Long, "VIPER: Visualization Pipeline for RNA-seq, a Snakemake workflow for efficient and complete RNA-seq analysis," *BMC Bioinformatics,* vol. 19, no. 1, pp. 135, Apr 12, 2018.

[3] S. Lewallen, and P. Courtright, "Epidemiology in practice: case-control studies," *Community Eye Health,* vol. 11, no. 28, pp. 57-8, 1998.

[4] B. Li, L. C. Tsoi, W. R. Swindell, J. E. Gudjonsson, T. Tejasvi, A. Johnston, J. Ding, P. E. Stuart, X. Xing, J. J. Kochkodan, J. J. Voorhees, H. M. Kang, R. P. Nair, G. R. Abecasis, and J. T. Elder, "Transcriptome analysis of psoriasis in a large case-control sample: RNA-seq provides insights into disease mechanisms," *J Invest Dermatol,* vol. 134, no. 7, pp. 1828-1838, Jul, 2014.

[5] G. XueKe, Z. Shuai, L. JunYu, L. LiMin, Z. LiJuan, and C. JinJie, "Lipidomics and RNA-Seq Study of Lipid Regulation in Aphis gossypii parasitized by Lysiphlebia japonica," *Sci Rep,* vol. 7, no. 1, pp. 1364, May 2, 2017.

[6] A. Mortazavi, B. A. Williams, K. McCue, L. Schaeffer, and B. Wold, "Mapping and quantifying mammalian transcriptomes by RNA-Seq," *Nat Methods,* vol. 5, no. 7, pp. 621-8, Jul, 2008.

[7] W. Huber, V. J. Carey, R. Gentleman, S. Anders, M. Carlson, B. S. Carvalho, H. C. Bravo, S. Davis, L. Gatto, T. Girke, R. Gottardo, F. Hahne, K. D. Hansen, R. A. Irizarry, M. Lawrence, M. I. Love, J. MacDonald, V. Obenchain, A. K. Oles, H. Pages, A. Reyes, P. Shannon, G. K. Smyth, D. Tenenbaum, L. Waldron, and M. Morgan, "Orchestrating high-throughput genomic analysis with Bioconductor," *Nat Methods,* vol. 12, no. 2, pp. 115-21, Feb, 2015.

[8] M. Pertea, D. Kim, G. M. Pertea, J. T. Leek, and S. L. Salzberg, "Transcript-level expression analysis of RNA-seq experiments with HISAT, StringTie and Ballgown," *Nat Protoc,* vol. 11, no. 9, pp. 1650-67, Sep, 2016.

[9] D. Kim, B. Langmead, and S. L. Salzberg, "HISAT: a fast spliced aligner with low memory requirements," *Nat Methods,* vol. 12, no. 4, pp. 357-60, Apr, 2015.

[10] M. Pertea, G. M. Pertea, C. M. Antonescu, T. C. Chang, J. T. Mendell, and S. L. Salzberg, "StringTie enables improved reconstruction of a transcriptome from RNA-seq reads," *Nat Biotechnol,* vol. 33, no. 3, pp. 290-5, Mar, 2015.

[11] Y. Kodama, M. Shumway, R. Leinonen, and C. International Nucleotide Sequence Database, "The Sequence Read Archive: explosive growth of sequencing data," *Nucleic Acids Res,* vol. 40, no. Database issue, pp. D54-6, Jan, 2012.

[12] C. N. Pang, Y. W. Lai, L. T. Campbell, S. C. Chen, D. A. Carter, and M. R. Wilkins, "Transcriptome and network analyses in Saccharomyces cerevisiae reveal that amphotericin B and lactoferrin synergy disrupt metal homeostasis and stress response," *Sci Rep,* vol. 7, pp. 40232, Jan 12, 2017.

[13] A. Conesa, P. Madrigal, S. Tarazona, D. Gomez-Cabrero, A. Cervera, A. McPherson, M. W. Szczesniak, D. J. Gaffney, L. L. Elo, X. Zhang, and A. Mortazavi, "A survey of best practices for RNA-seq data analysis," *Genome Biol,* vol. 17, pp. 13, Jan 26, 2016.

[14] H. B. TW, and T. Girke, "systemPipeR: NGS workflow and report generation environment," *BMC Bioinformatics,* vol. 17, pp. 388, Sep 20, 2016.

[15] P. Cliften, "Base Calling, Read Mapping, and Coverage Analysis," *Clinical Genomics*, pp. 91-107: Elsevier, 2015.

[16] B. Ewing, and P. Green, "Base-calling of automated sequencer traces using phred. II. Error probabilities," *Genome Res,* vol. 8, no. 3, pp. 186-94, Mar, 1998.

[17] A. Dobin, and T. R. Gingeras, "Mapping RNA-seq Reads with STAR," *Curr Protoc Bioinformatics,* vol. 51, pp. 11 14 1-19, Sep 3, 2015.

[18] A. C. Frazee, G. Pertea, A. E. Jaffe, B. Langmead, S. L. Salzberg, and J. T. Leek, "Ballgown bridges the gap between transcriptome assembly and expression analysis," *Nat Biotechnol,* vol. 33, no. 3, pp. 243-6, Mar, 2015.

[19] M. I. Love, W. Huber, and S. Anders, "Moderated estimation of fold change and dispersion for RNA-seq data with DESeq2," *Genome Biol,* vol. 15, no. 12, pp. 550, 2014.

[20] D. J. McCarthy, Y. Chen, and G. K. Smyth, "Differential expression analysis of multifactor RNA-Seq experiments with respect to biological variation," *Nucleic Acids Res,* vol. 40, no. 10, pp. 4288-97, May, 2012.

[21] M. D. Robinson, D. J. McCarthy, and G. K. Smyth, "edgeR: a Bioconductor package for differential expression analysis of digital gene expression data," *Bioinformatics,* vol. 26, no. 1, pp. 139-40, Jan 1, 2010.

[22] H. Wickham, *ggplot2: elegant graphics for data analysis*: Springer, 2016.

[23] T. Wei, and V. J. R. p. v. Simko, "corrplot: Visualization of a correlation matrix," vol. 230, no. 231, pp. 11, 2013.

[24] A. E. Berglund, E. A. Welsh, and S. A. Eschrich, "Characteristics and Validation Techniques for PCA-Based Gene-Expression Signatures," *Int J Genomics,* vol. 2017, pp. 2354564, 2017.

[25] S. Lê, J. Josse, and F. J. J. o. s. s. Husson, "FactoMineR: an R package for multivariate analysis," vol. 25, no. 1, pp. 1-18, 2008.

[26] R. J. R. p. v. Kolde, "Pheatmap: pretty heatmaps," vol. 61, 2012.

[27] G. Yu, L. G. Wang, Y. Han, and Q. Y. He, "clusterProfiler: an R package for comparing biological themes among gene clusters," *OMICS,* vol. 16, no. 5, pp. 284-7, May, 2012.

[28] M. Ashburner, C. A. Ball, J. A. Blake, D. Botstein, H. Butler, J. M. Cherry, A. P. Davis, K. Dolinski, S. S. Dwight, J. T. Eppig, M. A. Harris, D. P. Hill, L. Issel-Tarver, A. Kasarskis, S. Lewis, J. C. Matese, J. E. Richardson, M. Ringwald, G. M. Rubin, and G. Sherlock, "Gene ontology: tool for the unification of biology. The Gene Ontology Consortium," *Nat Genet,* vol. 25, no. 1, pp. 25-9, May, 2000.

[29] C. The Gene Ontology, "Expansion of the Gene Ontology knowledgebase and resources," *Nucleic Acids Res,* vol. 45, no. D1, pp. D331-D338, Jan 4, 2017.

[30] M. Kanehisa, and S. Goto, "KEGG: kyoto encyclopedia of genes and genomes," *Nucleic Acids Res,* vol. 28, no. 1, pp. 27-30, Jan 1, 2000.

[31] H. Ogata, S. Goto, K. Sato, W. Fujibuchi, H. Bono, and M. Kanehisa, "KEGG: Kyoto Encyclopedia of Genes and Genomes," *Nucleic Acids Res,* vol. 27, no. 1, pp. 29-34, Jan 1, 1999.

[32] W. Luo, and C. Brouwer, "Pathview: an R/Bioconductor package for pathway-based data integration and visualization," *Bioinformatics,* vol. 29, no. 14, pp. 1830-1, Jul 15, 2013.

[33] T. Barrett, S. E. Wilhite, P. Ledoux, C. Evangelista, I. F. Kim, M. Tomashevsky, K. A. Marshall, K. H. Phillippy, P. M. Sherman, M. Holko, A. Yefanov, H. Lee, N. Zhang, C. L. Robertson, N. Serova, S. Davis, and A. Soboleva, "NCBI GEO: archive for functional genomics data sets--update," *Nucleic Acids Res,* vol. 41, no. Database issue, pp. D991-5, Jan, 2013.

[34] R. Edgar, M. Domrachev, and A. E. Lash, "Gene Expression Omnibus: NCBI gene expression and hybridization array data repository," *Nucleic Acids Res,* vol. 30, no. 1, pp. 207-10, Jan 1, 2002.

[35] L. A. Martin, R. Ribas, N. Simigdala, E. Schuster, S. Pancholi, T. Tenev, P. Gellert, L. Buluwela, A. Harrod, A. Thornhill, J. Nikitorowicz-Buniak, A. Bhamra, M. O. Turgeon, G. Poulogiannis, Q. Gao, V. Martins, M. Hills, I. Garcia-Murillas, C. Fribbens, N. Patani, Z. Li, M. J. Sikora, N. Turner, W. Zwart, S. Oesterreich, J. Carroll, S. Ali, and M. Dowsett, "Discovery of naturally occurring ESR1 mutations in breast cancer cell lines modelling endocrine resistance," *Nat Commun,* vol. 8, no. 1, pp. 1865, Nov 30, 2017.

[36] S. K. Kim, S. Y. Kim, J. H. Kim, S. A. Roh, D. H. Cho, Y. S. Kim, and J. C. Kim, "A nineteen gene-based risk score classifier predicts prognosis of colorectal cancer patients," *Mol Oncol,* vol. 8, no. 8, pp. 1653-66, Dec, 2014.

[37] R. C. Team, "R: A language and environment for statistical computing," 2013.

[38] R. Gentleman, V. Carey, W. Huber, R. Irizarry, and S. Dudoit, *Bioinformatics and computational biology solutions using R and Bioconductor*: Springer Science & Business Media, 2006.

[39] R. C. Gentleman, V. J. Carey, D. M. Bates, B. Bolstad, M. Dettling, S. Dudoit, B. Ellis, L. Gautier, Y. Ge, and J. J. G. b. Gentry, "Bioconductor: open software development for computational biology and bioinformatics," vol. 5, no. 10, pp. R80, 2004.

[40] B. Grüning, R. Dale, A. Sjödin, B. A. Chapman, J. Rowe, C. H. Tomkins-Tinch, R. Valieris, J. Köster, and T. J. N. m. Bioconda,

"Bioconda: sustainable and comprehensive software distribution for the life sciences," vol. 15, no. 7, pp. 475, 2018.

[41] J. Wu, O. Anczukow, A. R. Krainer, M. Q. Zhang, and C. Zhang, "OLego: fast and sensitive mapping of spliced mRNA-Seq reads using small seeds," *Nucleic Acids Res,* vol. 41, no. 10, pp. 5149-63, May 1, 2013.

[42] A. Dobin, C. A. Davis, F. Schlesinger, J. Drenkow, C. Zaleski, S. Jha, P. Batut, M. Chaisson, and T. R. Gingeras, "STAR: ultrafast universal RNA-seq aligner," *Bioinformatics,* vol. 29, no. 1, pp. 15-21, Jan 1, 2013.

[43] D. Kim, G. Pertea, C. Trapnell, H. Pimentel, R. Kelley, and S. L. Salzberg, "TopHat2: accurate alignment of transcriptomes in the presence of insertions, deletions and gene fusions," *Genome Biol,* vol. 14, no. 4, pp. R36, Apr 25, 2013.

[44] C. Trapnell, B. A. Williams, G. Pertea, A. Mortazavi, G. Kwan, M. J. van Baren, S. L. Salzberg, B. J. Wold, and L. Pachter, "Transcript assembly and quantification by RNA-Seq reveals unannotated transcripts and isoform switching during cell differentiation," *Nat Biotechnol,* vol. 28, no. 5, pp. 511-5, May, 2010.

[45] A. I. Tomescu, A. Kuosmanen, R. Rizzi, and V. Makinen, "A novel min-cost flow method for estimating transcript expression with RNA-Seq," *BMC Bioinformatics,* vol. 14 Suppl 5, pp. S15, 2013.

[46] A. C. Frazee, G. Pertea, A. E. Jaffe, B. Langmead, S. L. Salzberg, and J. T. Leek, "Flexible analysis of transcriptome assemblies with Ballgown," 2014.

[47] J. Goecks, A. Nekrutenko, J. Taylor, and T. Galaxy, "Galaxy: a comprehensive approach for supporting accessible, reproducible, and transparent computational research in the life sciences," *Genome Biol,* vol. 11, no. 8, pp. R86, 2010.

[48] D. Hull, K. Wolstencroft, R. Stevens, C. Goble, M. R. Pocock, P. Li, and T. Oinn, "Taverna: a tool for building and running workflows of services," *Nucleic Acids Res,* vol. 34, no. Web Server issue, pp. W729-32, Jul 1, 2006.

[49] J. Koster, and S. Rahmann, "Snakemake-a scalable bioinformatics workflow engine," *Bioinformatics,* vol. 34, no. 20, pp. 3600, Oct 15, 2018.

[50] P. Di Tommaso, M. Chatzou, E. W. Floden, P. P. Barja, E. Palumbo, and C. J. N. b. Notredame, "Nextflow enables reproducible computational workflows," vol. 35, no. 4, pp. 316, 2017.

[51] C. Guzman, and I. D'Orso, "CIPHER: a flexible and extensive workflow platform for integrative next-generation sequencing data analysis and genomic regulatory element prediction," *BMC Bioinformatics,* vol. 18, no. 1, pp. 363, Aug 8, 2017.

[52] R. V. J. E. j. Guimera, "bcbio-nextgen: Automated, distributed next-gen sequencing pipeline," vol. 17, no. B, pp. 30, 2011.

[53] M. Wolfien, C. Rimmbach, U. Schmitz, J. J. Jung, S. Krebs, G. Steinhoff, R. David, and O. Wolkenhauer, "TRAPLINE: a standardized and automated pipeline for RNA sequencing data analysis, evaluation and annotation," *BMC Bioinformatics,* vol. 17, pp. 21, Jan 6, 2016.

[54] D. Wang, "hppRNA-a Snakemake-based handy parameter-free pipeline for RNA-Seq analysis of numerous samples," *Brief Bioinform,* vol. 19, no. 4, pp. 622-626, Jul 20, 2018.

[55] S. Zhao, L. Xi, J. Quan, H. Xi, Y. Zhang, D. von Schack, M. Vincent, and B. J. B. g. Zhang, "QuickRNASeq lifts large-scale RNA-seq data analyses to the next level of automation and interactive visualization," vol. 17, no. 1, pp. 39, 2016.

[56] D. Gaidatzis, A. Lerch, F. Hahne, and M. B. Stadler, "QuasR: quantification and annotation of short reads in R," *Bioinformatics,* vol. 31, no. 7, pp. 1130-2, Apr 1, 2015.

[57] M. J. Steinbaugh, L. Pantano, R. D. Kirchner, V. Barrera, B. A. Chapman, M. E. Piper, M. Mistry, R. S. Khetani, K. D. Rutherford, and O. J. F. Hofmann, "bcbioRNASeq: R package for bcbio RNA-seq analysis," vol. 6, 2018.

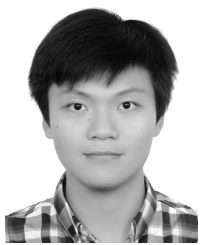

**Kuan-Hao Chao** is an undergraduate student in the Department of Electrical Engineering at the National Taiwan University (NTU) and will be graduating in 2020 with a BS in Electrical Engineering. His research projects mainly focus on developing bioinformatics online systems and Bioconductor R package.

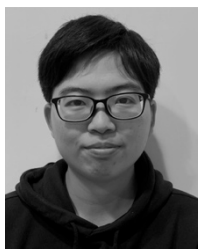

**Yi-Wen Hsiao** obtained her Master degree from the Industrial and Commercial Biotechnology program, Newcastle University. She is a research assistant in the Bioinformatics and Biostatistics Core Lab, Center of Genomic and Precision Medicine, National Taiwan University.

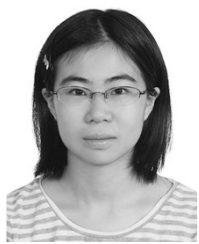

**Yi-Fang Lee** is currently EECS master of engineering student at UC Berkeley. With interests in both engineering and biology, she earned her bachelor's degree in Electrical Engineering and master of science degree in Bioinformatics at National Taiwan University (NTU), Taiwan. Her research projects mainly focus on developing bioinformatics databases and online systems.

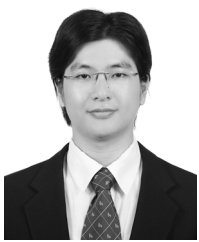

**Chien-Yueh Lee** received the BSc degree in electrical engineering from National Taipei University of Technology, Taiwan, and the MSc and Ph.D. degrees in biomedical electronics and bioinformatics from National Taiwan University, Taiwan. He is a postdoctoral researcher in Perelman School of Medicine, University of Pennsylvania, USA. His research interests include bioinformatics, computational biology, genomics, big data analysis, and database.

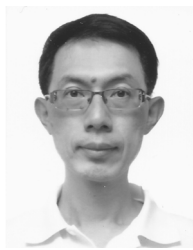

**Liang-Chuan Lai** earned his PhD from University of Illinois at Urbana-Champaign in 2005. He is an Associate Professor in Institute of Biotechnology, National Taiwan University. His major research interests are using genomic approaches to explore the molecular mechanism of carcinogenesis.

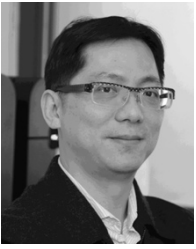
**Mong-Hsun Tsai** received his PhD from Institute of Public Health, National Yang-Ming University in 2001. He is a Professor in Institute of Biotechnology, National Taiwan University. His interests are bioinformatics, cell biology, microarray and radiation biology.

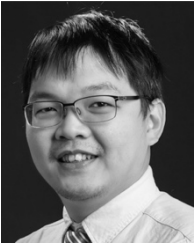
**Tzu-Pin Lu** obtained his PhD from the Graduate Institute of Biomedical Electronics and Bioinformatics, National Taiwan University. He is an Associate Professor in the Institute of Epidemiology and Preventive Medicine, College of Public Health, National Taiwan University. His major research interests includes bioinformatics and computational biology.

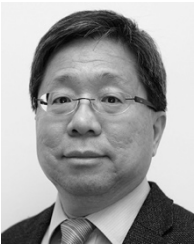
**Eric Y. Chuang** is a Professor in Graduate Institute of Biomedical Electronics and Bioinformatics, National Taiwan University. He earned his ScD from Harvard University in cancer biology in 1997. His major research area are biochip, bioinformatics and cancer biology.